 \documentclass[preprint2]{emulateapj}
 \usepackage{graphicx}

\shorttitle{Dust in the reionization era: ALMA observations of a $z$=8.38 Galaxy}
\shortauthors{Laporte et al.}

\begin{document}


\title{Dust in the Reionization Era: ALMA Observations of a $z=$8.38 Gravitationally-Lensed Galaxy}


\author{N. Laporte\altaffilmark{1,3,4},
R. S. Ellis\altaffilmark{1,2},
F. Boone\altaffilmark{6,7},
F. E. Bauer\altaffilmark{3,4,5},
D. Qu\'enard\altaffilmark{8},
G. Roberts-Borsani\altaffilmark{1},
R. Pell\'o\altaffilmark{6,7}, 
I. P\'erez-Fournon\altaffilmark{9,10}, and
A. Streblyanska\altaffilmark{9,10}
}


\altaffiltext{1}{Department of Physics and Astronomy, University College London, Gower Street, London WC1E 6BT, UK}
\altaffiltext{2}{European Southern Observatory (ESO), Karl-Schwarzschild-Strasse 2, 85748 Garching, Germany}
\altaffiltext{3}{Instituto de Astrof{\'{\i}}sica and Centro de Astroingenier{\'{\i}}a, Facultad de F{\'{i}}sica, Pontificia Universidad Cat{\'{o}}lica de Chile, Casilla 306, Santiago 22, Chile} 
\altaffiltext{4}{Millennium Institute of Astrophysics (MAS), Nuncio Monse{\~{n}}or S{\'{o}}tero Sanz 100, Providencia, Santiago, Chile} 
\altaffiltext{5}{Space Science Institute, 4750 Walnut Street, Suite 205, Boulder, Colorado 80301} 
\altaffiltext{6}{Universit\'e de Toulouse; UPS-OMP; IRAP; Toulouse, France}
\altaffiltext{7}{CNRS; IRAP; 14, avenue Edouard Belin, F-31400 Toulouse, France}
\altaffiltext{8}{School of Physics and Astronomy Queen Mary, University of London 327 Mile End Road, London, E1 4NS}
\altaffiltext{9}{ Instituto de Astrof\'isica de Canarias, C/V\'ia L\'actea, s/n, E-38205 San Crist\'obal de La Laguna, Tenerife, Spain}
\altaffiltext{10}{Universidad de La Laguna, Dpto. Astrof\'isica, E-38206 La Laguna, Tenerife, Spain}


\begin{abstract}
We report on the detailed analysis of a gravitationally-lensed Y-band dropout, A2744\_YD4, selected from deep \textit{Hubble Space Telescope} imaging in the \textit{Frontier Field} cluster Abell 2744. Band 7 observations with the Atacama Large Millimeter Array (ALMA) indicate the proximate detection of a significant 1mm continuum flux suggesting the presence of dust for a star-forming galaxy with a photometric redshift of $z\simeq8$. Deep X-SHOOTER spectra confirms the high redshift identity of A2744\_YD4 via the detection of Lyman $\alpha$ emission at a redshift $z$=8.38. The association with the ALMA detection is confirmed by the presence of [OIII] 88$\mu$m emission at the same redshift.  Although both emission features are only significant at the 4 $\sigma$ level, we argue their joint detection and the  positional coincidence with a high redshift dropout in the HST images  confirms the physical association. Analysis of the available photometric data and the modest gravitational magnification ($\mu\simeq2$) indicates A2744\_YD4 has a stellar mass of $\sim$ 2$\times$10$^9$ M$_{\odot}$, a star formation rate of $\sim20$ M$_{\odot}$/yr and a  dust mass of $\sim$6$\times$10$^{6}$ M$_{\odot}$. We discuss the implications of the formation of such a dust mass only $\simeq$200 Myr after the onset of cosmic reionisation. 
\end{abstract}



\section{Introduction}

The first billion years of cosmic history represents the final frontier in assembling a coherent physical picture of early galaxy formation and considerable progress has been enabled through observations from space-based telescopes and ground-based optical and near-infrared spectrographs. Early progress in the \textit{Hubble} Ultra Deep and the CANDELS fields (\citealt{2013ApJ...763L...7E}, \citealt{2015ApJ...803...34B}, \citealt{2015ApJ...810...71F}) has been complemented by surveys through lensing clusters (\citealt{2012ApJS..199...25P}), an approach culminating in \textit{Hubble Space Telescope}'s (HST) flagship program, the \textit{Frontier Fields} (FF) \citep{2016arXiv160506567L}. By harnessing the magnification of a foreground massive cluster, galaxies of more representative luminosities can be probed (e.g. \citealt{2014A&A...562L...8L}, \citealt{2016ApJ...820...98L}, \citealt{2015ApJ...800...84C}, \citealt{2015ApJ...814...69A}). Collectively, the blank field and cluster surveys have located several hundred star-forming galaxies in the redshift range $6<z<10$ corresponding to the era when it is thought hydrogen was photo-ionized (\citealt{2016A&A...594A..13P}, \citealt{2015ApJ...802L..19R}). In addition to the population demographics analyzed through photometric data from \textit{HST} and the \textit{Spitzer Space Telescope}, spectroscopic diagnostics are being gathered to gauge the nature of their stellar populations and their capability for releasing the necessary number of ionizing photons (for a recent review see \citealt{2016ARA&A..54..761S}). 


The completion of the Atacama Large Millimeter Array (ALMA) brings the possibility of measuring the dust content of these early systems. Dust is likely produced by the first supernovae and timing its formation would measure the extent of previous star formation. Moreover, dust can affect many of the key physical properties derived from  photometric data. While early ALMA observations focused on distant $z\simeq$6 massive ultra-luminous galaxies, targeting the more representative lower mass systems in the reionization era brought into view by gravitational lensing is an interesting approach. An exciting early result was the detection of a significant dust mass in a $z\simeq$7.5 galaxy whose rest-frame UV colors indicated little or no reddening \citep{2015Natur.519..327W}.  


The present paper is concerned with the follow-up and analysis of an ALMA continuum detection in the FF cluster Abell 2744 close to a Y-band dropout, A2744\_YD4, at a photometric redshift of $z\simeq$8.4. 
In Section \ref{sec.obs} we introduce the ALMA 1mm continuum detection and its possible association with A2744\_YD4  and justify  a photometric redshift of $z\simeq$8 for this galaxy. In Section \ref{sec.spectro} we analyze deep X-Shooter spectra which confirms the redshift via the detection of Ly$\alpha$ emission supported by O [III] 88 $\mu$m emission within the Band 7 ALMA data. We deduce the physical properties and dust mass of A2744\_YD4 in Section \ref{sec.properties} and discuss the implications for early dust formation in \ref{sec.discussion}. Throughout this paper, we use a concordance cosmology ($\Omega_M = 0.3$, $\Omega_{\Lambda} = 0.7$ and $H_0 = 70$ km sec$^{-1}$ Mpc$^{-1}$ and all magnitudes are quoted in the AB system \citep{1983ApJ...266..713O}.


\section{Imaging Data}
\label{sec.obs}
Here we describe the ALMA data in which a high-$z$ candidate is detected at 0.84mm, and the public imaging data used to constrain its Spectral Energy Distribution (SED).

\subsection{Deep ALMA band 7 observations}
\begin{figure*}
\centering
\includegraphics[width=17cm] {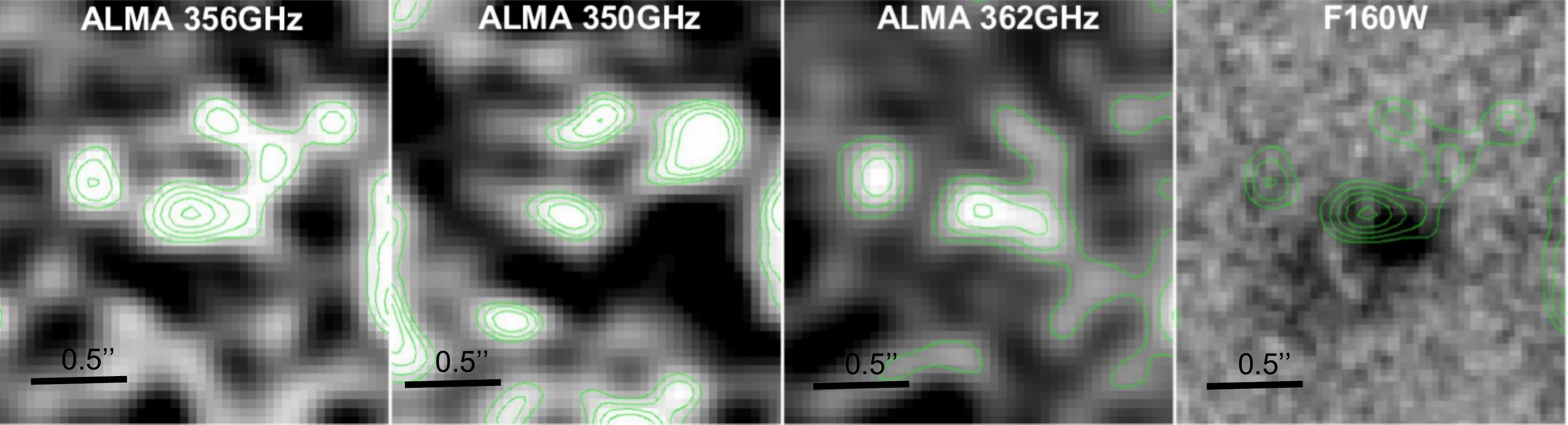}
\caption{\label{Fig.alma.map} ALMA Band 7 continuum detection for A2744\_YD4. (\textit{Left}): map combining all frequency channels, (\textit{Center pair}) independent maps for two equal frequency ranges. Contours are shown at 1, 2, 3, 4 and 5$\sigma$ adopting a noise level from an area of 0.5$\times$0.5. arcmin. (\textit{Right}) HST F160W image with combined ALMA image contours overplotted.}
\end{figure*}

A deep ALMA Band 7 map (ID 2015.1.00594, PI: Laporte) of the FF cluster Abell 2744 centered at 0.84mm ($f_c$=356 GHz) was observed on July 2016 during 2.5hrs. The data was reduced using the CASA pipeline \citep{2007ASPC..376..127M} with a natural weighting and a pixel size of 0.04''. Figure \ref{Fig.alma.map} reveals a source with greater than $4.0\sigma$ significance with a peak flux of 9.9$\pm$2.3 $\times$10$^{-5}$ Jy/beam (uncorrected for magnification). The uncertainty and significance level was computed from the rms measured across a representative $\approx$2$\times$2 arcmin field. The signal is seen within two independent frequency ranges (center panels in Figure \ref{Fig.alma.map}) and the significance level is comparable to that claimed for Watson et al's $z\sim$7.5 lensed system although its observed flux is six times fainter.  Taking into account the different magnification factors (see later), the intrinsic (lensing-corrected) peak fluxes are similar at $\approx$ 5$\times$10$^{-5}$ Jy.  Dividing the exposure into two independent halves, the significances of 3.2 and 3.4 are consistent with that of the total exposure.

To identify the likely source, we examined the final version of the reduced HST data of Abell 2744 (ACS and WFC3) acquired between November 2013 and July 2014 as part of the Frontier Fields program (ID: 13495 -- PI : Lotz), combining this with archival data from previous campaigns (ID : 11689 ; PI : Dupke  -- ID: 13386 ; PI : Rodney). Although there is some structure in the ALMA detection, it lies close to the source A2744\_YD4 (F160W=26.3) at RA= 00:14:24.9,  DEC=$-$30:22:56.1(2000) first identified by \citet{2014ApJ...795...93Z}.  Correcting for an astrometric offset between HST positions and astrometry measured by the Gaia telescope \citep{2016A&A...595A...2G}, we deduce a small physical offset of $\simeq$0.2 arcsec between the ALMA detection and the HST image.



\subsection{Other Imaging Data}

\begin{figure}
\includegraphics[width=7.5cm]{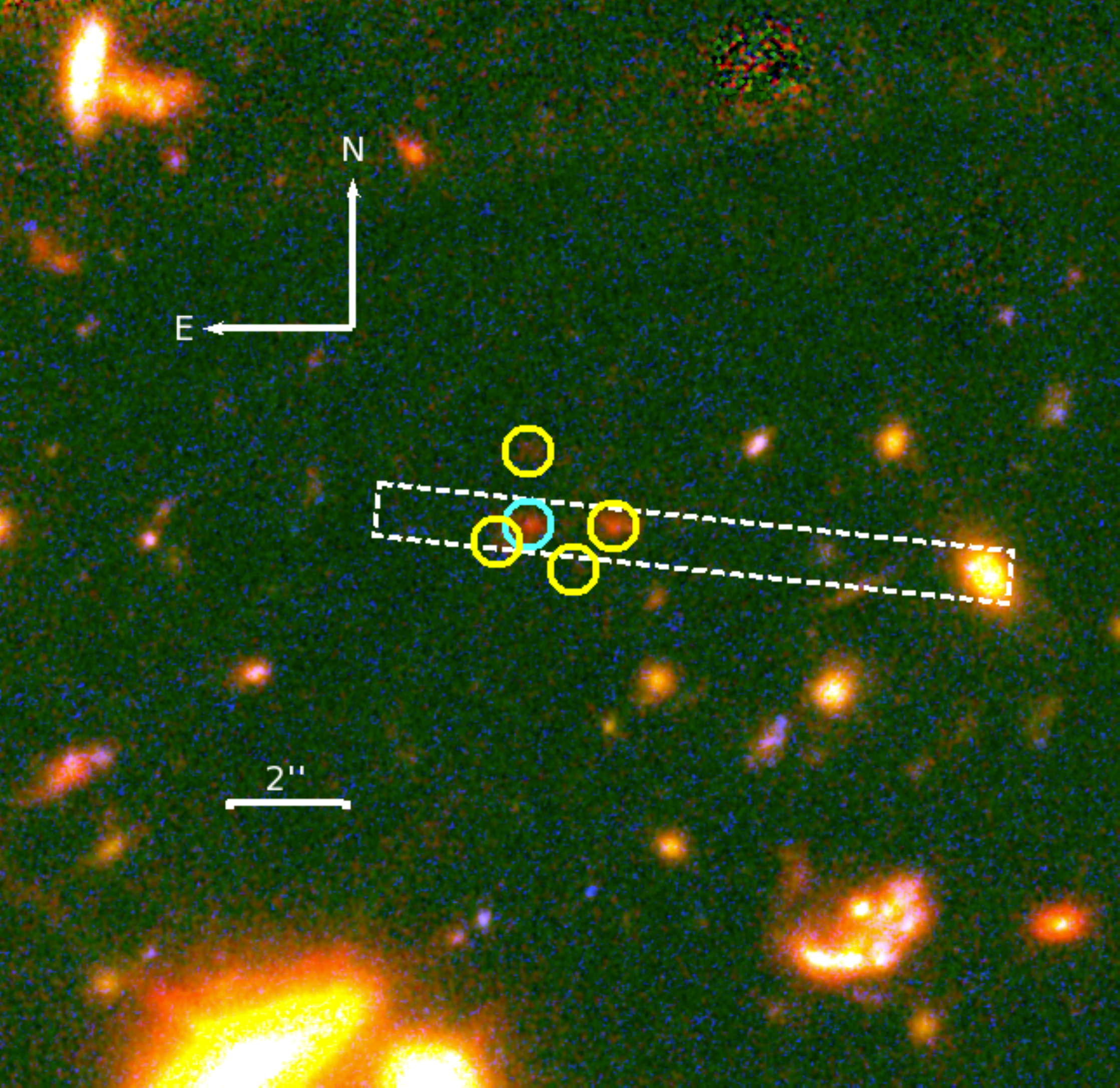}
\caption{\label{Fig.slit} Position of the ALMA band 7 detected high redshift galaxy A2744\_YD4 (blue) respect to other group members (yellow) suggested by \citet{2014ApJ...795...93Z}. The X-shooter slit orientation is shown with the dashed white line. Although one other member of the group was targeted in the exposure, no confirming features were found in the data. }
\end{figure}

Deep $K_s$ data is also available from a  29.3 hrs HAWK-I image taken between October and December 2013 (092.A-0472 -- PI: Brammer) which reaches a 5$\sigma$ depth of 26.0. Spitzer IRAC data obtained in channel 1 ($\lambda_c \sim$3.6$\mu$m) and 2 ($\lambda_c \sim$4.5$\mu$m) with 5$\sigma$ depths of 25.5 and 25.0 respectively carried out under DDT program (ID: 90257, PI: T. Soifer).  We extracted the HST photometry on PSF-matched data using \textit{SExtractor} \citep{1996A&AS..117..393B} v2.19.5 in double image mode using the F160W map for the primary detection (Figure \ref{Fig.alma.map}). To derive the total flux, we applied an aperture correction based on the F160W MAG\_AUTO measure (see e.g. \citealt{2006ApJ...653...53B}). The noise level was determined using several 0.2 arcsec radius apertures distributed around the source. The total $K_s$ magnitude of 26.45 $\pm$ 0.33 was obtained using a 0.6 arcsec diameter aperture applying the correction estimated in \citet{2016ApJS..226....6B}. The uncertainty was estimated following a similar procedure to that adopted for the HST data.  The Spitzer data was reduced as described in Laporte et al (2014) using corrected Basic Calibrated Data (cBCD) and the standard reduction software MOPEX to process, drizzle and combine all data into a final mosaic.  As shown in Figure \ref{Fig.slit}, four other galaxies are close to A2744\_YD4, but only the other source within the X-shooter slit is comparably bright to A2744\_YD4. We used GALFIT \citep{2002AJ....124..266P} to deblend the two sources and to measure their IRAC fluxes. We fitted both IRAC ch1 and ch2 images assuming fixed positions, those measured from the F160W image.
 Our photometry of A2744\_YD4 is consistent with that published previously by the AstroDeep team (\citealt{2016A&A...590A..30M}, \citealt{2014ApJ...795...93Z} and \citealt{2015ApJ...800...84C}).

\subsection{SED Fitting}
\label{sec.phot_and_sed}

We used several SED fitting codes to estimate the photometric redshift of A2744\_YD4 and hence its implied association with the ALMA detection. In each case we fit all the available photometric data (HST-ACS, HST-WFC3, VLT HAWKI, Spitzer). 

Firstly, we used an updated version of \textit{Hyperz} \citep{2000A&A...363..476B} with a template library drawn from \citet{2003MNRAS.344.1000B}, \citet{2001ApJ...556..562C}, \citet{1980ApJS...43..393C}, 
and \citet{1999ApJS..123....3L} including nebular emission lines as described by \citet{2009A&A...502..423S}. We permitted a range in redshift (0 $< z <$ 10) and extinction (0 $ < A_v <$ 3) and found the best solution at $z_{phot}$=8.42$^{+0.09}_{-0.32}$ ($\chi^2\sim$1), with no acceptable solution at lower redshift. Restricting the redshift range to 0$< z < 3$ and increasing the extinction interval to (0 $ < A_v <$ 10), we found a low redshift solution at $z_{phot}^{low-z}$=2.17$^{+0.03}_{-0.08}$ but with a significantly worse $\chi^2$ $\sim$9. 

We also made use of the Easy and Accurate Zphot from Yale (\textit{EAZY}; \citealt{brammer08}) software. The SED fits adopted the standard SED templates from \textit{EAZY}, as well as those from the Galaxy Evolutionary Synthesis Models (GALEV; \citealt{kotulla09})  including nebular emission lines as described by \citet{anders03}. Adopting a large redshift range (0 $< z <$ 10) with no prior assumptions on the extinction, the best-fit has $z_{phot}$=8.38$^{+0.13}_{-0.11}$ in excellent agreement with that from \textit{Hyperz}. 

In summary, the photometry strongly supports a $z>8$ solution. A low $z$ solution is unlikely given the F814W - F125W $>$ 3mag break as well as the low statistical likelihood. 

\begin{figure}
\includegraphics[width=9.5cm]{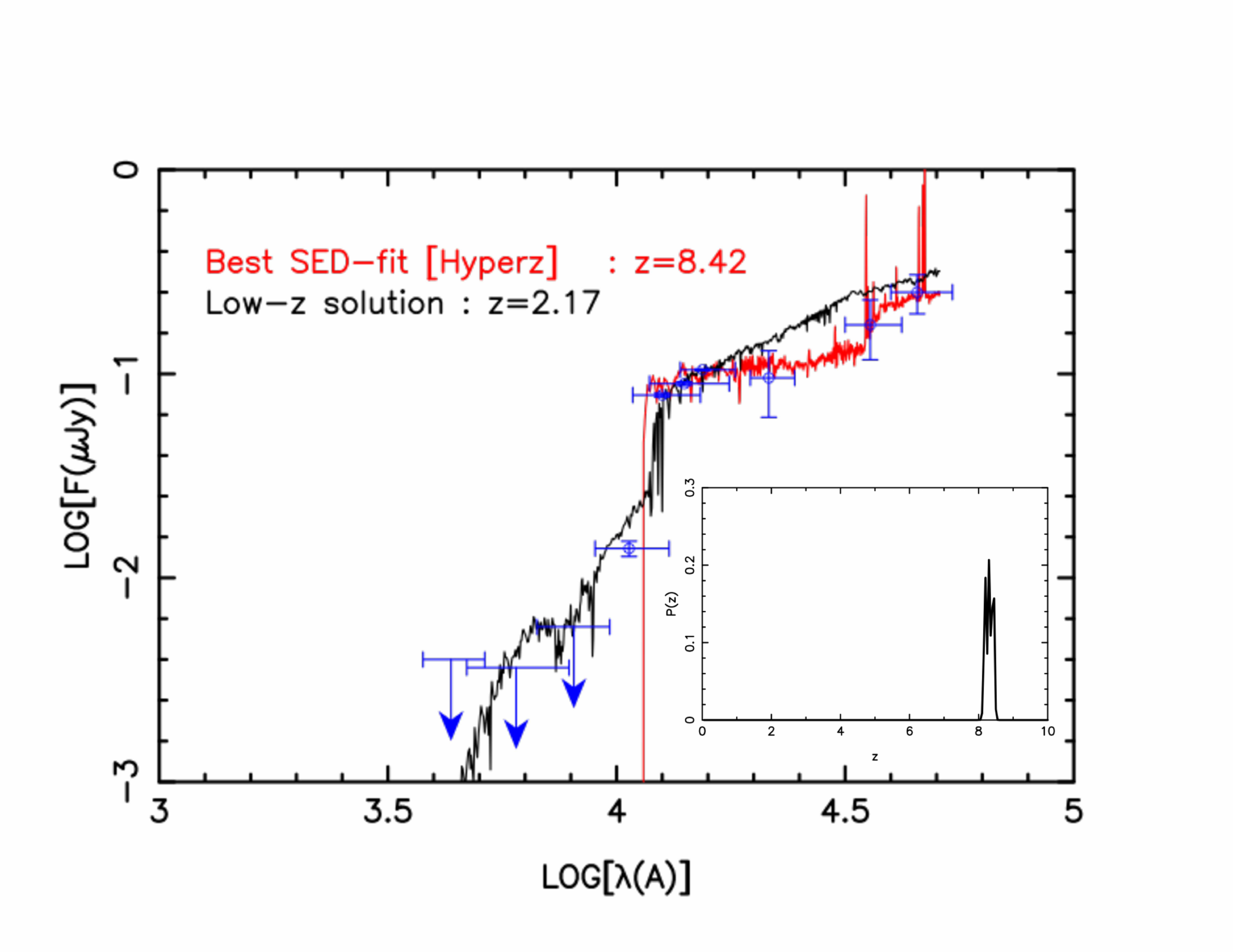}
\caption{\label{Fig.SED} Spectral Energy Distribution of A2744\_YD4. The red curve shows the best fitting SED found by \textit{Hyperz} with $z_{phot}$=8.42$^{+0.09}_{-0.32}$. The black curve shows a forced low redshift solution derived when only a redshift interval from 0 to 3 is permitted. This has a likelihood $>$20 times lower. The inner panel displays the redshift probability distribution.}
\end{figure}

\section{Spectroscopic Follow-up}
\label{sec.spectro}

\subsection{X-Shooter Observations}

Given the importance of confirming the presence of dust emission beyond $z\simeq$8, we undertook a spectroscopic campaign using X-Shooter/VLT (ID: 298.A-5012 -- PI: Ellis). Between 24-27 November 2016, we secured 7.5 hours on-source integration with excellent seeing ($\approx$0.6 arcsec ).  We used a 5 arcsec dither to improve the sky subtraction and aligned the slit so that a brighter nearby source could verify the blind offset (see Fig. \ref{Fig.slit}). The data was reduced using  v2.8 of the ESO Reflex software 
combined with X-Shooter pipeline recipes v2.8.4. 

\begin{figure*}
\centering
\includegraphics[width=9.cm]{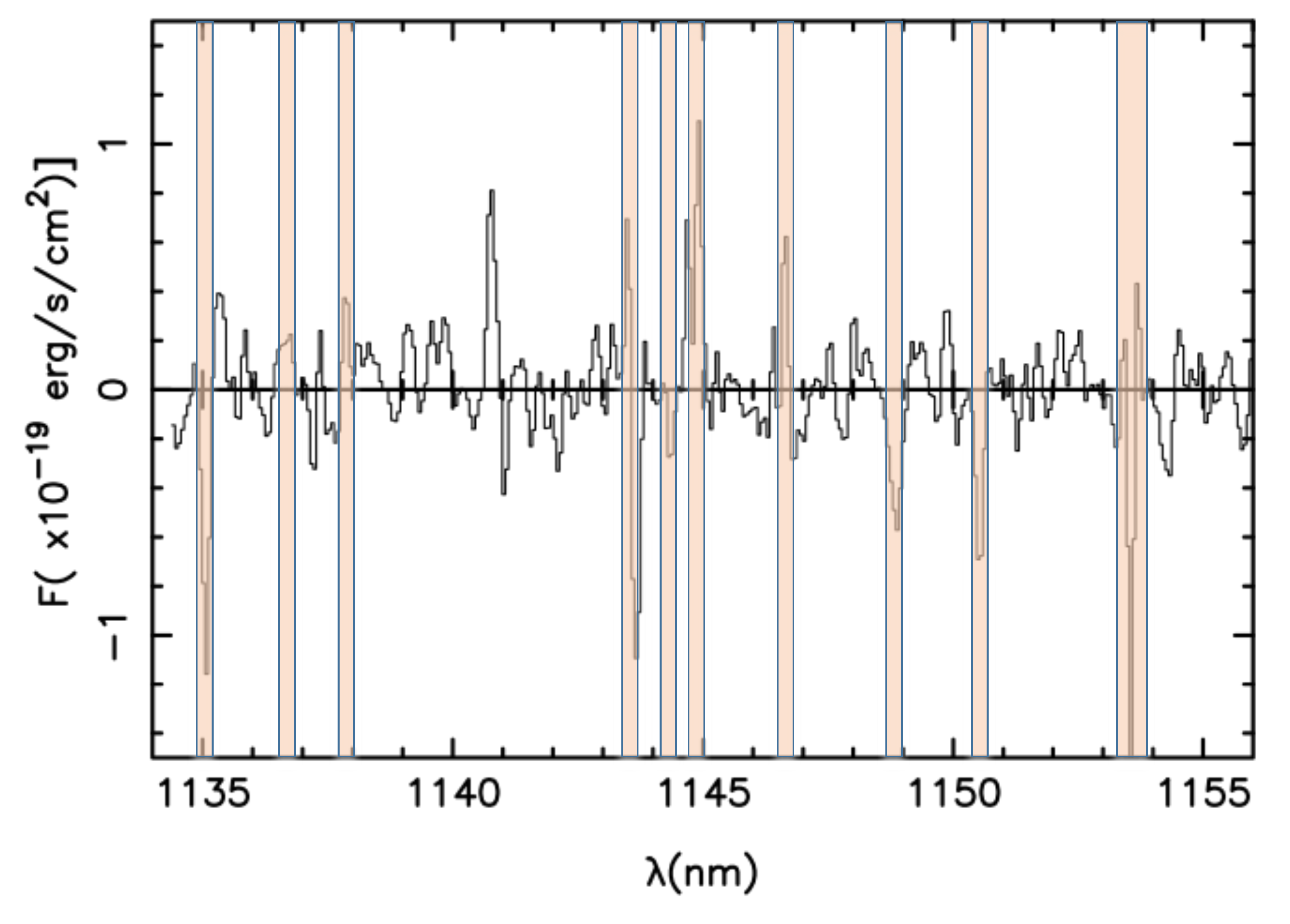} 
\includegraphics[width=6.cm]{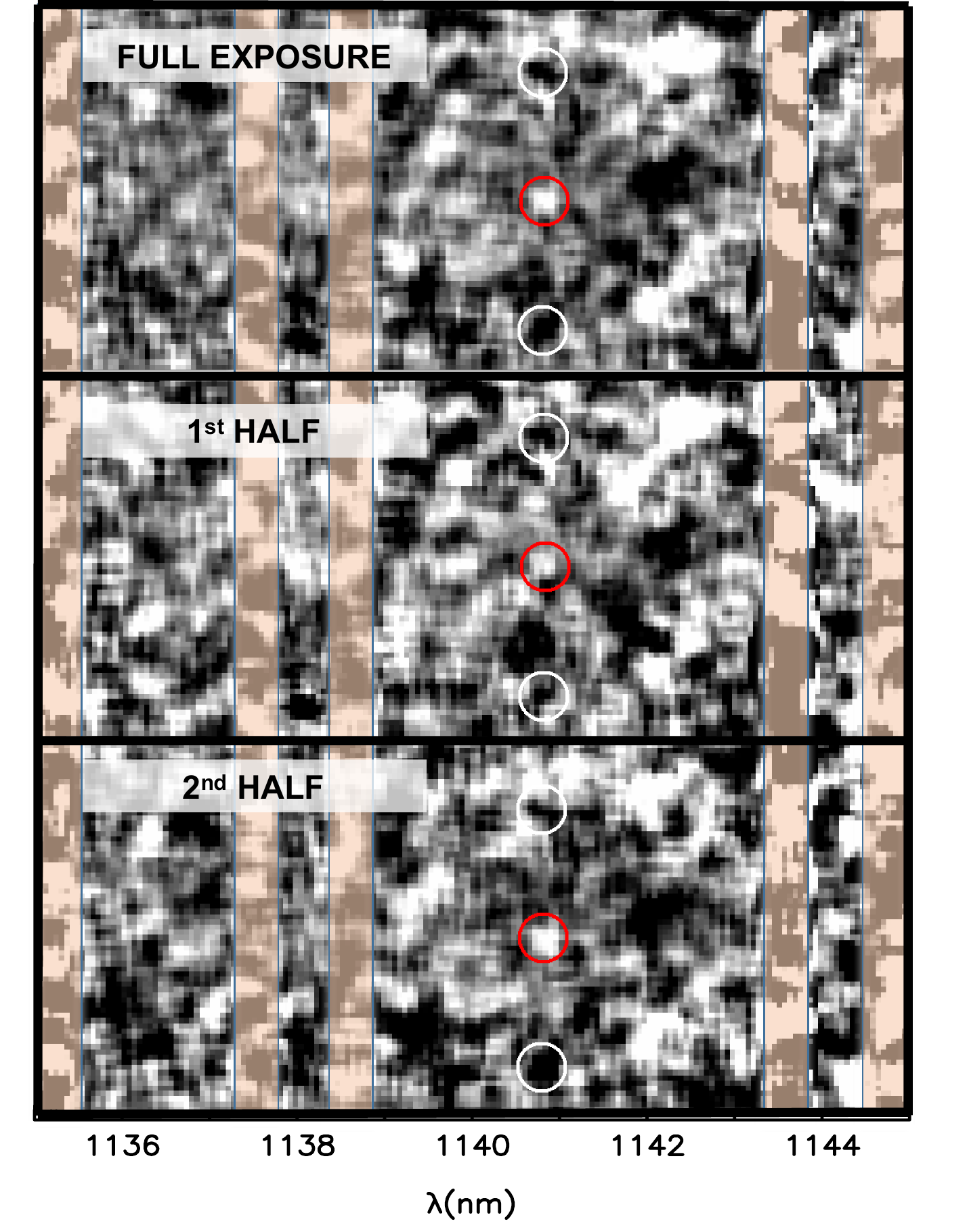}

\caption{\label{fig.Lya} 
(\textit{Left}) Extracted 1D spectrum with OH night sky contamination indicated in orange . (\textit{Right}) 2D spectra separated into (top) the total exposure (7.5hrs on source), (center) first half of the total exposure, (bottom) latter half of the total exposure.}
\end{figure*}

We visually inspected all 3 arms of the X-Shooter (UVB, VIS, NIR) spectrum and identified one emission line at $\lambda$=11408.4 \AA\ with an integrated flux of $f$=1.82$\pm$0.46$\times$10$^{-18}$erg s$^{-1}$ cm$^{-2}$.  By measuring the rms at adjacent wavelengths we measure the significance as $\approx$4.0$\sigma$.  We checked the reliability of the line by confirming its presence on two independent spectral subsets spanning half the total exposure time (Figure \ref{fig.Lya}).  These half exposures show the line with significances of 2.7 and 3.0, consistent with that of the total exposure. No further emission lines of comparable significance were found. We explore two interpretations of this line at 11408 \AA\ . It is either (1)  one component of the [OII] doublet at a redshift $z\simeq$2.06, or (2) Ly$\alpha$ at $z$=8.38.

For (1), depending on which component of the [OII]$\lambda$3727,3730 doublet is detected, we expect a second line at either 11416.9 \AA\ or 11399.8 \AA\ . 
 No such emission is detected above the 1$\sigma$ flux limit of 4.6$\times$10$^{-19}$ cgs. This would imply a flux ratio for the two components of $\approx$3.95  (2.02) at 1(2) $\sigma$, greater than the range of 0.35-1.5 from theoretical studies (e.g. \citealt{2006MNRAS.366L...6P}).

 For (2) although the line is somewhat narrow for Ly$\alpha$ (rest-frame width $\approx$ 20km s$^{-1}$), its equivalent width deduced from the line flux and the F125W photometry is 10.7 $\pm$ 2.7 \AA\ , consistent with the range seen in other $z>7$ spectroscopically-confirmed sources \citep{2017MNRAS.464..469S}. We detect no flux above the noise level at the expected position of either the CIV and [OIII] doublets at this redshift.  However, at the expected position of the CIII] doublet,  we notice a very marginal ($\approx$2$\sigma$) feature at $\lambda$=17914.7\AA\ (7.5$\pm$0.35$\times$10$^{-19}$erg s$^{-1}$ cm$^{-2}$) seen on two individual sub-exposures. If this is CIII]$\lambda$1907\AA\ (normally the brighter component ) at $z_{CIII}$=8.396, the Ly-$\alpha$ offset of 338$\pm$3 km s$^{-1}$ would be similar to that for a $z$=7.73 galaxy \citep{2017MNRAS.464..469S}. The other component would be fainter than 5.0$\times$10$^{-19}$erg s$^{-1}$ cm$^{-2}$ consistent with theoretical studies of this doublet (e.g. \citealt{2004ApJ...605..784R}).
 

Previous spectroscopy of A2774\_YD4 was undertaken by the GLASS survey \citep{2016ApJ...818...38S} who place a 1$\sigma$ upper limit on any Lyman-$\alpha$ detection at 4.4$\times$10$^{-18}$ erg s$^{-1}$ cm$^{-2}$, $\approx$2.4 times above our X-Shooter detection. 


\subsection{ALMA Observations}

\begin{figure}[h]
\centering
\includegraphics[width=9cm] {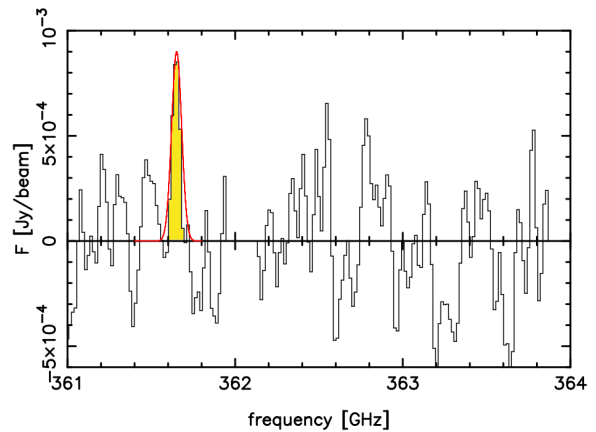}
\caption{ \label{fig.alma.line} The ALMA [OIII]88$\mu$m spectrum with a resolution of 25km s$^{-1}$. The best Gaussian profile is over-plotted in red at the central frequency corresponding to a redshift of $z$=8.382. The derived line width is $\approx$43km s$^{-1}$}.   
\end{figure}

Only a few far infrared emission lines lines are detectable for sources in the reionization era (see eg. \citealt{2013ASPC..476...23C}). Only the [OIII] 88$\mu$m line at the  $z$=8.38 redshift of Ly-$\alpha$ would be seen in the frequency range covered by our ALMA observations. Given the recent detection of this line in a $z\simeq$7.2 Lyman alpha emitter \citep{2016Sci...352.1559I}, we examined our ALMA data for such a possibility. Searching our data in frequency space, we find a 4.0$\sigma$ narrow emission line offset by 0.35 arcsec from the astrometric position of A2744\_YD4 at a frequency of 361.641 GHz.  Dividing the exposure time in half, the line is detected with independent significances of 2.8 and 3.2, consistent with that of the total exposure.
Assuming this line is [OIII] 88$\mu$m, the redshift would be $z$=8.382 (see Figure \ref{fig.alma.line}), leading to a Lyman-$\alpha$ velocity shift of $\sim$70km s$^{-1}$ in good agreement with that observed in a $z\sim$7.2 galaxy \citep{2016Sci...352.1559I}.  Fitting the emission line with a Gaussian profile we derive a modest FWHM=49.8 $\pm$ 4.2 km s$^{-1}$ implying an intrinsic width of $\simeq$43 km s$^{-1}$. The emission line luminosity is estimated at 1.40$\pm$0.35$\times$10$^8$L$_{\odot}$ without correction for magnification, which is $\approx$ 7 times fainter than that detected in Inoue et al's $z=$7.2 source. The peak line flux of A2744\_YD4 is consistent with that computed from simulations in \citet{2014ApJ...780L..18I} (see their Fig. 3).  Compared to the aforementioned lower mass source at $z=$7.2, the narrower line width is perhaps surprising but may indicate its formation outside the body of the galaxy as inferred from the offset and recent simulations \citep{2017MNRAS.466.1648K}.



\section{Physical Properties}
One of the main objectives of this study is to utilize the spectroscopic redshift as well as the ALMA band 7 detection to estimate accurate physical properties for A27744\_YD4, and particularly to constrain the dust mass for an early star forming galaxy.
\label{sec.properties}
\subsection{Magnification}

Estimating the magnification is critical to determine the intrinsic properties of any lensed source. Several teams have provided mass models for each of the six clusters. Moreover a web tool\footnote{https:\/\/archive.stsci.edu\/prepds\/frontier\/lensmodels\/\#magcalc} enables us to estimate the magnification for Abell2744\_YD4 from parametric high resolution models, i.e. version 3.1 of the CATS model \citep{2014MNRAS.444..268R}, version 3 of \citet{2014ApJ...797...48J}, version 3 of \citet{2011MNRAS.417..333M} and version 3 of GLAFIC \citep{2015ApJ...799...12I}. We took the average value with error bars corresponding to the standard deviation: $\mu$=1.8$\pm$0.3.

\subsection{The Star Formation Rate and Stellar and Dust Masses}

The detection of dust emission in a $z$=8.38 galaxy provides an unique opportunity to evaluate the production of dust, presumably from early supernovae in the first few 100 Myr since
reionization began. The key measures are the dust and stellar masses and the likely average past star formation rate.

We first estimate some physical properties based on the ALMA continuum detection using a simple modified black body SED with the dust temperatures ranging from 35 to 55K and the dust emissivity fixed at $\beta$$\sim$2. We found a total FIR luminosity ranging from 7.1 to 18.2 $\times$10$^{10}$ L$_{\odot}$ and a dust mass ranging from 1.8 to 10.4$\times$10$^{6}$ M$_{\odot}$. These values are corrected for magnification and CMB heating.

We also ran an updated version of MAGPHYS \citep{2008MNRAS.388.1595D} adapted for high-$z$ galaxies \citep{2015ApJ...806..110D}. The code estimates the properties of A2744\_YD4 in two steps. First, it generates a library of model SEDs at the redshift of our source ($z=$8.38) in our 11 bands (7 from HST, the deep HAWKI K$_s$ band,  the two first IRAC channels and a synthetic filter of the ALMA band 7) for a wide range of variables including the star formation rate (SFR) and dust content. We generated a total of about 9 million models, including $\approx$25,000 IR dust emission models.  MAGPHYS then derives the likelihood distribution of each physical parameter by comparing the observed SED with all the models in the library. 
In this way we derived the following properties: SFR=20.4$^{+17.6}_{-9.5}$ M$_{\odot}$ yr$^{-1}$;  a stellar mass M$_{\star}$=(1.97$^{+1.45}_{-0.66}$)$\times$10$^{9}$ M$_{\odot}$; a dust mass M$_{dust}$=(5.5$^{+19.6}_{-1.7}$)$\times$10$^6$ M$_{\odot}$, and an extinction A$_v$=0.74$^{+0.17}_{-0.48}$ with a dust temperature ranging from 37 to 63 K. The error bars are refer to 1$\sigma$ uncertainties. These values estimated with MAGPHYS using the full SED are consistent with those deduced solely from the ALMA continuum flux. 

Although the uncertainties in these physical properties are large, our target is similar to the lensed source A1689-zD1 at $z\simeq$7.5 studied by \citet{2015Natur.519..327W}. \citealt{2015Natur.519..327W} reports M$_{\star}$=1.7$^{+0.7}_{-0.5}$$\times$10$^{9}$ M$_{\odot}$ and SFRs 9$^{+4}_{-2}$ M$_{\odot}$/yr. Their specific star formation rates (sSFR) are thus similar (1.04$^{+10.2}_{-0.21}$$\times$10$^{-8}$ yr$^{-1}$ for A2744\_YD4 and 0.6$^{+1.1}_{-0.3}$$\times$10$^{-8}$ yr$^{-1}$ for A1689-zD1) implying a mean lifetime for a constant SFR of $\approx$ 100 Myr. However, A1689-zD1 has a significantly larger dust mass of M$_{dust}$=4 $^{+4}_{-2} \times$ 10$^{7}$ M$_{\odot}$. Possibly this is a consequence of continuous star formation over a longer period together with more advanced chemical enrichment.


\section{Discussion}
\label{sec.discussion}

\citet{2014ApJ...795...93Z} identified A2744\_YD4 as one member of a group composed of 5 galaxies  with similar colors and photometric redshifts. Although there is no spectroscopic information for the others members, conceivably this group is contained within a  single ionized bubble, which may explain the detection of the Lyman-$\alpha$ emission in what is currently considered to be an epoch when the IGM is fairly neutral\citep{2015ApJ...802L..19R}. The putative group is contained within an  area of 1.7 arcsec radius or 8.1 kpc. (Figure \ref{Fig.slit}). A second group member was included on the X-shooter slit; although no emission was seen, it is over 1 mag fainter in continuum luminosity than A2744\_YD4.

We finally turn to the implications of dust emission at such a high redshift. According to the most recent analyses of the history of cosmic reionization (\citealt{2015ApJ...802L..19R}, \citealt{2016A&A...594A..13P}, \citealt{2016MNRAS.459.2342M}), significant star formation began at $z\simeq$10-12, about 200 Myr before the epoch at which A2744\_YD4 is being observed. The dust output and rate of early supernovae is of course highly uncertain but, for a past average star formation rate of 20 $M_{\odot}$ yr$^{-1}$, assuming a popular stellar initial mass function (e.g. \citealt{1955ApJ...121..161S}
) with a high mass power law slope of $\simeq-$7/3, we expect $\simeq$0.2\% of newly-born stars to exceed 8 $M_{\odot}$ and produce Type II SNe. Assuming each SN produces around 0.5 M$_{\odot}$ of dust in its core \citep{2015ApJ...800...50M}, over 200 Myr this would yield around 4$\times$ 10$^6$ M$_{\odot}$ of dust in apparent agreement with the observations.  However, this would not account for any dust lost to the system given typical velocities of ejection could be 10$^{2-3}$ km s$^{-1}$.

These speculations are as far as we can proceed given the current uncertainties. The most important conclusion is that ALMA clearly has the potential to detect dust emission within the heart of the reionization era and thus further measures of this kind, in conjuction with spectroscopic verification both using ALMA and soon the James Webb Space Telescope, offers the exciting prospect of tracing the early star formation and onset of chemical enrichment out to redshifts of 10 and beyond.


\acknowledgments{
We acknowledge discussions with Dan Stark, Rob Ivison,  Harley Katz and Jason Spyromilio. NL and RSE acknowledge support from the ERC Advanced Grant FP7/669253. D. Q. acknowledges support from a STFC Rutherford Grant (ST/M004139/2). FEB acknowledges CONICYT-Chile (Basal-CATA PFB-06/2007, FONDECYT Regular 1141218), the Ministry of Economy, Development, and Tourism's Millennium Science Initiative through grant IC120009. IPF acknowledges support from the grant ESP2015-65597-C4-4-R of the Spanish
Ministerio de Economía y Competitividad (MINECO). This study is based on observations made at ESO's Paranal Observatory under programme ID 298.A-5012, on the following ALMA data: ADS/JAO.ALMA \#[2015.1.00594], on observations obtained with the NASA/ESA Hubble Space Telescope, retrieved from the Mikulski Archive for Space Telescopes (MAST) at the Space Telescope Science Institute (STScI). ALMA is a partnership of ESO (representing its member states), NSF (USA) and NINS (Japan), together with NRC (Canada), NSC and ASIAA (Taiwan), and KASI (Republic of Korea), in cooperation with the Republic of Chile. The Joint ALMA Observatory is operated by ESO, AUI/NRAO and NAOJ. STScI is operated by the Association of Universities for Research in Astronomy, Inc. under NASA contract NAS 5-26555. We used gravitational lensing models produced by PIs Bradac, Natarajan \& Kneib (CATS), Merten \& Zitrin, Sharon, and Williams, and the GLAFIC and Diego groups. This lens modeling was partially funded by the HST Frontier Fields program conducted by STScI. STScI is operated by the Association of Universities for Research in Astronomy, Inc. under NASA contract NAS 5-26555. }




\clearpage




\end{document}